\documentclass[reprint,amsmath,amssymb,aps,prb]{revtex4-2}
\usepackage{import}
\usepackage{xcolor}
\usepackage{tabularx}
\newcolumntype{Y}{>{\centering\arraybackslash}X}
\usepackage{svg}
\usepackage[newcommands]{ragged2e}
\usepackage{graphicx}
\usepackage{dcolumn}
\usepackage{bm}
\usepackage{chemformula} % Formula subscripts using \ch{}
\usepackage[T1]{fontenc} % Use modern font encodings
\usepackage{comment}
\usepackage{physics}
\usepackage[utf8]{inputenc}
\usepackage[colorlinks=true,linkcolor=blue,filecolor=magenta,urlcolor=blue,citecolor=blue,pdfstartview=FitH]{hyperref}
\usepackage[justification=raggedright]{caption}
\usepackage{subcaption}
\usepackage{float}
\graphicspath{{figures/}}
\usepackage{blindtext}
\usepackage{array}
\newcolumntype{P}[1]{>{\centering\arraybackslash}p{#1}}
\usepackage{subfiles}
\usepackage{amsmath}
\usepackage{amsfonts}
\usepackage{amssymb}
\usepackage{multirow}
\usepackage{xr}

\usepackage{tabularray}
\UseTblrLibrary{amsmath}
\makeatletter

\definecolor{clr1}{rgb}{0,0,0}
\definecolor{clr2}{rgb}{0.0,0.35,0.0}
\newcommand{\figref}[1]{Fig.~\ref{#1}}
\newcommand{\secref}[1]{Sec.~\ref{#1}}
\newcommand{\eqaref}[1]{Eq.~\eqref{#1}}
\newcommand{\apref}[1]{Appendix~\ref{#1}}

\newcommand{\beq}{\begin{equation}}
	\newcommand{\eeq}{\end{equation}}
\newcommand{\ie}{i.e.,~}
\newcommand{\clr}[1]{\color{black}#1\color{black}}

\newcommand{\bse}{\begin{subequations}}
	\newcommand{\ese}{\end{subequations}}
\newcommand{\bea}{\begin{eqnarray}}
	\newcommand{\eea}{\end{eqnarray}}
\newcommand{\bem}{\begin{displaymath}}
	\newcommand{\eem}{\end{displaymath}}
\newcommand{\bmat}{\begin{bmatrix}}
	\newcommand{\ebmat}{\end{bmatrix}}

\newcommand{\bs}{\boldsymbol}
\newcommand{\bc}{\begin{center}}
	\newcommand{\ec}{\end{center}}

\begin{document}	
	\title{Magnetically modulated superconductor-graphene-superconductor (SGS)  Josephson junctions and their tunability}	
	\author{Partha Sarathi Banerjee}
	\affiliation{Department of Physics,
		Indian Institute of Technology Delhi,
		Hauz Khas, New Delhi 110016, India}	
	\author{Rahul Marathe}
	\affiliation{Department of Physics,
		Indian Institute of Technology Delhi,
		Hauz Khas, New Delhi 110016, India}	
	\author{Sankalpa Ghosh}
	\affiliation{Department of Physics,
		Indian Institute of Technology Delhi,
		Hauz Khas, New Delhi 110016, India}	
%	\date{April 10, 2024}
\date{\today}

		\begin{abstract}
		
		Graphene-based Josephson junctions \cite{Heersche2007, Du2008, Popinciuc2012, Mizuno2013, Efetov2016, bretheau2017, Borzenets2016}  played an important role in various quantum devices from their inception. 
		Magnetic tunnel junctions  or vertical devices \cite{Cobas2012, Meng2013,Chen2013,Huang2024} were also made out of graphene by exposing the graphene layer to localised pattern of strong magnetic field  created by hard ferromagnetic material.
		By combining the essence of these different  methods for constructing graphene based junctions, in this work we propose that the temperature-dependent Josephson current in such junctions can be tuned 
		by exposing the graphene regions to a combination of highly localised non-uniform magnetic field, dubbed as magnetic barrier, and spatially modulated gate voltage. Within the framework of Dirac-Bogoliubov-de-Gennes (DBDG)  theory,  we show by explicit calculation 
		that in such magnetically modulated Josephson Junctions,  the band structure of graphene gets significantly altered, which results in the change of the  Andreev reflections in such junctions. This leads to a significant modulation of the Josephson current. 
		We numerically evaluated the Josephson current as a function of the strength of the magnetic barrier and the gate voltage and discussed the practical consequences of such controlling of Josephson currents. 
			\end{abstract}

\maketitle
	
\section{Introduction}
The dissipation-less super-current, namely the Josephson current (JC) that flows through Josephson junctions (JJ) \cite{Josephson1962, Waldram1970}  played a significant role in quantum technologies such as superconducting qubit devices \cite{arute2019quantum,Christianson2008,wallraff2004strong, Castellanos2007,you2002scalable}, sensing small magnetic fields \cite{bal2012ultrasensitive,vettoliere2015long}, parametric amplifiers \cite{frattini20173,frattini2018optimizing}, single photon detection \cite{walsh2021josephson}, etc., to name a few.  The JC decreases with increasing temperature, which is the primary tuning parameter for such JJs for  given materials, and vanishes at the critical temperature \cite{Lim1970, Yoshida2004, Kivioja2005}. The JJs  have been made with a thin insulating layer \cite{Cyster2021, Makhlin2001, Wendin2007}, metal \cite{Cakir2003, Golubov2004, kulik1969}, two-dimensional-electron-gas\cite{nguyen1990inas, Nitta1992}, ferromagnet\cite{Salehi2010}, and in recent times using mono and bilayer graphene where two closely spaced superconducting electrodes are placed on a graphene(G) sheet to make an SGS type \cite{Titov2006, Li2016, bretheau2017} of JJ due to proximity induced superconductivity \cite{Shailos2007, Lee2018, BlackSchaffer2008}. 
As first pointed out by Beenakker \cite{Beenakker2006}\clr{, } compared to other weak links, the  Josephson effect in SGS junctions can be attributed to  both  specular Andreev reflection (SAR) 
%as compared to 
\clr{ and } conventional retro Andreev reflection (RAR) \cite{andreev1964thermal, kulik1969}, 
due to the peculiar quasiparticle dispersion in graphene \cite{Castro2009, novoselov2005two, zhang2005experimental, Mccann2013}. Tunability of JC in such junctions using different means is therefore desirable for wider applicability of such devices.

%When an electron reaches the interface from the weak link. It is reflected as a hole and a Cooper pair travels through the S region. This process, known as Andreev reflection etermines the conductivity of the JJ below the superconducting gap. I. Due to linear dispersion and the massless nature of the relativistic graphene electrons the %inear bands, the massless low-energy excitations in graphene are described by the Dirac equation. Due to high electronic mobility, external field tunable chemical potential and sensitive thermal response, graphene is a favorable choice to make the JJs than semiconductor or metals \cite{walsh2021josephson, Borzenets2016}.  

\begin{figure*}
	\centering
	\includegraphics[width=1.0\linewidth]{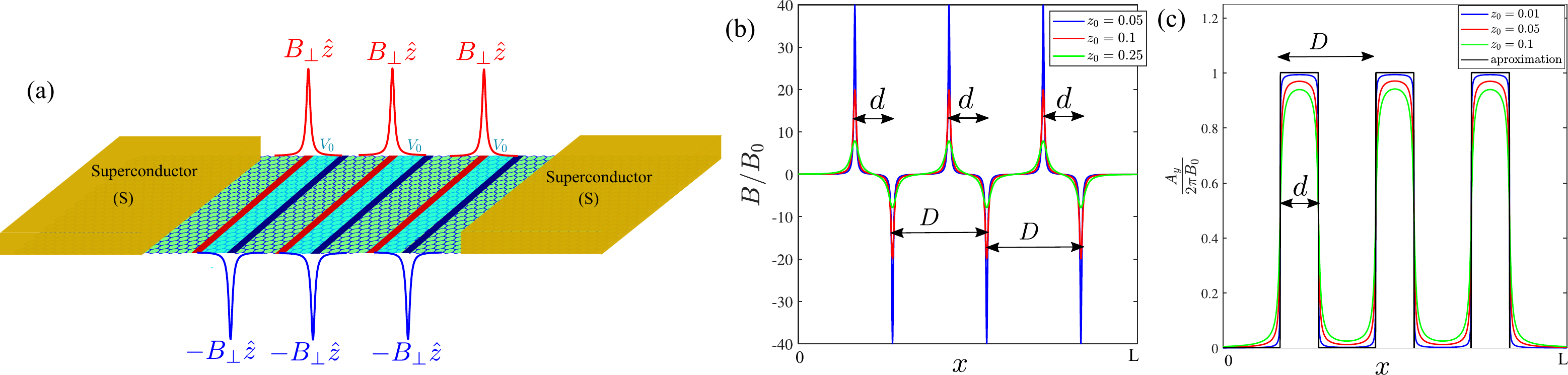}
	\caption{\justifying In (a) the schematic depiction of the SGS type JJ with periodic electric and magnetic field. in the graphene region is shown. In the regions between the red and blue magnetic stripes a non zero electric field ($V_0$) is also considered. In (b) we compare the perpendicular magnetic field profile as seen by the massless Dirac fermions for different values of $z_0$. This can be done by changing the distance between the ferromagnetic stripe and the graphene sheet. The corresponding magnetic vector potential is shown in (c). As $z_0$ is reduced the magnetic field barriers become more close to a perfectly rectangular barrier.}
	\label{schem}
\end{figure*}

In this paper, we show that the temperature-dependent JC in such SGS junctions can be made more tunable for possible device applications by exposing the graphene region to a regularly spaced highly localised magnetic field typically dubbed in the literature as magnetic barriers \cite{Martino2007, DeMartino2007, Oroszlany2008, Ramezani2009, Ghosh2009, Masir2010,Le2012}.  In  our proposed model\clr{, } such SGS junctions are exposed to magnetic barriers that can be created by putting a ferromagnetic strip\clr{e } on top of the surface of graphene region \cite{Mancoff1995, Kubrak1999}. In experiments, such metallic strip\clr{es } have been made by \ch{NdFeB} \cite{Mancoff1995}, \ch{Co} \cite{Kubrak1999}. They emit a strong magnetic field to create this highly inhomogeneous magnetic field which breaks the time reversal symmetry in the graphene region explicitly. Using these stripes, a magnetic vector potential barrier can be created where two of such stripes  produce a strong magnetic fields of equal magnitude, but in mutually opposite directions \cite{You1995} which is transverse to the plane of the graphene layer as depicted in \figref{schem} (a). In graphene, using a fully scalable photolithographic process, \ch{Co} based ferromagnetic layer has been deposited to make a magnetic tunnel junction \cite{Cobas2012, Meng2013,Chen2013,Huang2024}. \color{black}  This makes the synthesis of  such magnetically modulated SGS junctions a realistic possibility.

Because of the ultra-relativistic dispersion of the charge carriers in monolayer graphene, the strength of such magnetic barriers can be additionally tuned by an electrostatic barrier \cite{Peters2010, Stander2009, Sharma2011, Chen2016}. These potential barriers can be created by putting a gate voltage to create a p-n junction in graphene \cite{Williams2007, Cheianov2006, Park2008}. Array of such barriers can be viewed as a realisation of magnetic Kronig-Penny model \cite{Ramezani2009,Masir2010,Le2012}. They consequently modify the band structure,  thereby impacting  Andreev reflection \cite{andreev1964thermal} in such relativistic Josephson junction. 
This, in turn, changes the JC flowing through such JJs. The combined electrostatic and magnetic barrier is schematically depicted in \figref{schem} (a). 

JJs with high transition temperature ($T_C$) and higher superconducting energy gaps provide advantages for qubit applications involving graphene, as they are operable for higher temperature range. Superconductors with \ch{Al} \cite{Heersche2007, Du2008, Miao2009}, \ch{W} \cite{Shailos2007}, \ch{Pt/Ta} \cite{Ojeda2009} and \ch{Pb_{1-x}In_x} \cite{Jeong2011} as superconducting electrodes has been used in experiments to make Josephson junctions. 
Keeping that in mind, in our calculations, we consider the superconducting gap for \ch{Pb_{0.93}In_{0.07}} electrode at $1.1$ meV and the transition temperature at $7.0$ K \cite{Jeong2011}. It may be noted that even though \ch{Pb} ($T_C=7.19$K) and \ch{In} ($T_C=3.4$K) are type-I superconductors, their alloy becomes a type-II superconductor. The transition temperature also depends on their composition \cite{Evetts1970}.

The rest of the paper is organized as follows. 
The formalism and the DBdG equations in the three types of regions are discussed in \secref{sec2}. In \secref{Snell} we discuss the  electronic analogue of the Snell's law in the magnetically modulated graphene region and their impact on the Andreev refelction. In \secref{boundary} we calculate the transfer matrices from the boundary value conditions. The method to calculate the $\varepsilon-\phi$  relation for these systems is shown in \secref{disp}. The Josephson current for both SAR and RAR regimes, and how the Josephson current can be modulated by the geometry and the magnetic barrier is discussed in \secref{curr}.

\section{Superconductor-Graphene-Superconductor Junctions in the Presence of Electric and Magnetic Barrier in the Graphene Region} \label{sec2}

The system under consideration is a two-dimensional monolayer graphene (MLG) in the $x$-$y$ plane, which  is covered with a superconducting electrode in two regions ($x<0$ and $x>L$)  (see \figref{schem} (a)). 
The regions under superconducting electrode become two-dimensional superconductor\clr{s } due to the proximity effect \cite{bretheau2017, Lee2018, Beenakker2006}.
Due to the difference between the phases of the two superconductors ($\phi$), a dissipation-less current flows in these systems which is periodic in $\phi$ \cite{Josephson1962}. 
%\color{blue}
The non-superconducting graphene region in between ($0 \le x \le L$), is exposed to ferromagnetic stripe\clr{s } with thickness $d$ and height $h$, that are deposited on top of the two-dimensional graphene sheet with magnetization parallel \cite{Cobas2012} to the plane of the MLG.

If the graphene layer is situated at distance $z_0$ below the stripe then the magnetic field in the perpendicular direction can be written by \cite{Matulis1994}
 $B_z \hat{z}= \hat{z} K(x,z_0)$ where,
\begin{equation}
	K(x,z_0) = B_0 \left[ \frac{z_0 d}{x^2+z_0^2}- \frac{z_0 d}{(x-d)^2+z_0^2}\right]
\end{equation}
The profile magnetic vector potential corresponding to this magnetic field takes form of a barrier, that shifts the momentum of the charge carriers. A series of $N$ such magnetic vector potential barriers can be created, where the successive barriers are separated by a distance $D$. The resulting magnetic field becomes,
\begin{equation}
	B_z \hat{z}= \hat{z} \sum_{m=0}^{N-1} K\left[x- (m+1) D +d , z_0\right].
\end{equation}

For $N$ such barriers in graphene region with uniform spacing between them, $L=N D + (D-d)$, where both $d$ and $D$ ( defined in the figure)  varied in a given situation to achieve wider tunability. The magnetic field profile is shown in \figref{schem}(b) and the corresponding magnetic vector potential in Landau gauge, $\mathbf{A}=\hat{y} A_y$ is shown in \figref{schem} (c). For a typical value of $B_0=0.1$ T \cite{Majumdar1996, Fujita2010, Matulis1994}, which depends on the magnetization($M_0$), height ($h$) and
thickness ($d$) of the stripe as  $B_0=M_0 h/d$, outside the magnetic barrier regime 
the magnetic field is several orders of magnitude less than the critical
field of \ch{Pb_{0.93}In_{0.07}} electrode that we are considering
\cite{Jeong2011}. A typical value for $B_{0}$ can be inferred from the fact that  magnetic fields upto the order of $\approx 1$T \cite{Huang2024} has been used to create Co based magnetic tunnel junctions in graphene. 
In an experimental setup, the magnetic potential profile can be changed by changing the parameters $z_0$ and $d$.  In the limit $z_0 \rightarrow 0$, the magnetic field profiles become delta functions, and the corresponding magnetic vector potentials approaches a perfect rectangular barrier. 
We perform our further calculation in this limit, and the approximated vector potentials are shown by the black curve in \figref{schem} (c). 

To proceed further we write the magnetic vector potentials in Landau gauge as,

   \begin{equation}
 	\mathbf{A}(x)= \begin{cases}
 		\hat{y} B l_m &\text{in the barrier region } (n D -d) <x <nD, \\
 		0		&\text{in the G region } (n-1)D < x < (nD -d),\\
 		0	&\text{in the S region}~x<0 \text{ and }x>L.
 	\end{cases} \label{vpot}
 \end{equation}

Here, $n=1,2,...N$, $l_m=\sqrt{\frac{\hbar c}{eB}}$ is the magnetic length. As suggested in the introduction, for $0 \le x \le L$, we additionally consider a series of electrostatic potential barriers coupled with the magnetic barriers with the same width $d$ and height $V(V>0)$ \cite{Sharma2011}. 
These electrostatic potentials can be created by putting a series of electric gates \cite{Peters2010, Stander2009, Chen2016}. An additional constant electrostatic potential $-U_0$, whose value can be adjusted by an external gate potential or by doping \cite{Beenakker2006}
in the superconducting region , is also introduced to offset the difference in electrostatic potential in the superconducting region and the graphene region. The resulting scalar electrostatic potential is given by 

\begin{equation}
	V(x)=\begin{cases}
		-U_0&\text{in the S region } x<0 \text{ and }x>L,\\
		0&\text{in the G region } (n-1)D < x < (nD -d),\\
		V&\text{in the barrier region } (n D -d) <x <nD.
	\end{cases}
\end{equation}

The dynamics of charge carriers, namely the electron and hole excitations, 
 in  such SGS junction can be described by Dirac-Bogoliubov-De-Gennes (DBdG) equation \cite{de1966superconductivity,Beenakker2006, Titov2006,Maiti2007, Akhmerov2007}, and can be written as 
\begin{equation}\label{DBDG}
	\begin{bmatrix}
		\mathcal{H} - \mu && \Delta(T) \\
		\Delta^*(T)&& \mu - \mathcal{T}\mathcal{H}\mathcal{T}^{-1}
	\end{bmatrix}\begin{bmatrix}
		\Psi_e\\
		\Psi_h
	\end{bmatrix}= \varepsilon \begin{bmatrix}
		\Psi_e\\
		\Psi_h
	\end{bmatrix} 
\end{equation}
Here $\mathcal{H}$ is the $(4\times4)$ Dirac Hamiltonian for the charge carriers in graphene under effective mass approximation, namely massless Dirac fermions, in the presence of the electrostatic potential $V(x)$ and magnetic field $\mathbf{B}$, which acts on two sub-lattice and two valley degrees of freedom for such charge carriers. $v_F$ is the Fermi velocity, $\Delta(T)$ is the temperature dependent superconducting pair potential which couples the time reversed electron and hole states. For our present analysis, we consider s-wave pair potential such that 
\begin{equation}
	\Delta(T)=\begin{cases}
		\Delta_0(T) e^{i \phi_1} &\text{ at, } x<0\\
		0  &\text{ at, } 0<x<L \\
		\Delta_0(T) e^{i \phi_2} &\text{ at, } x>L.\\
	\end{cases}
\end{equation}
This can be generalised for more complicated pairing potential. The temperature dependence of such superconducting pair potential is given by \cite{Tinkham2004},
\begin{equation}
	\Delta_0 (T)= \Delta_0 (T=0) \sqrt{1- \left(\frac{T}{T_C}\right)^2},
\end{equation}
where critical temperature of a superconductor is given by $T_C$. 
 $\mathcal{T}$ is the time-reversal operator and $\varepsilon$ is the excitation energy measured relative to the Fermi energy $\mu$. $\Psi_e$ and $\Psi_h$ represent the electron and hole excitations. The time-reversal operator which interchanges the valleys is given by \cite{Beenakker2006, Titov2006, Suzuura2002},

\begin{equation}
	\mathcal{T}= \mqty[
	0 & \sigma_z\\
	\sigma_z & 0
	] \mathcal{C}= \mathcal{T}^{-1}.
\end{equation}

Here, $\mathcal{C}$ is the complex conjugation operator. As the pair-potential $\Delta(T)$  is same for both sub-lattice and valley degrees of freedom, in the presence of a magnetic field, it is more convenient to use a "valley-isotropic" basis for the Hamiltonian \cite{Akhmerov2007} by making a unitary transformation to both the Hamiltonian and the time-reversal operator $H = \mathcal{U} \mathcal{H} \mathcal{U}^{\dagger}$ and $T = \mathcal{U} \mathcal{T} \mathcal{U}^{\dagger}$ with $\mathcal{U}=\frac{1}{2}(\tau_0+\tau_z)\otimes\sigma_0+ \frac{1}{2}(\tau_0-\tau_z)\otimes \sigma_x$. $\sigma_i$ and $\tau_i$ are the Pauli spin matrices which act on the sub-lattice and valley degrees of freedom respectively. Also, $\tau_0$ and $\sigma_0$ represent $2\times2$ unit matrices. Upon these transformations the Hamiltonian now becomes,
\begin{align}
	H &= v_F \mqty[
	\left[(\mathbf{p} + e \mathbf{A}) \cdot \bs{\sigma}+ \frac{V(x)}{v_F} \right] & 0\\
	0 &	\left[(\mathbf{p} + e \mathbf{A}) \cdot \bs{\sigma}+ \frac{V(x)}{v_F} \right]
	] \notag \\
	&=v_F \tau_0 \otimes \left[(\mathbf{p} + e \mathbf{A}) \vdot \bm{\sigma}+ \frac{V(x)}{v_F} \right].	\label{ham}
\end{align}

The time-reversal operator now becomes,

\begin{equation}
	T= \mqty[
	0	&	i \sigma_y\\
	-i \sigma_y		&	0	
	] \mathcal{C} = -(\tau_y\otimes \sigma_y)  \mathcal{C}. \label{trev}
\end{equation}

Using the Hamiltonian and the time-reversal operator from the \eqaref{ham} and \eqref{trev} in the Dirac Bogoliubov De Gennes (DBdG) equation \eqaref{DBDG} in the presence of external potential $V(x)$ and magnetic vector potential $\mathbf{A}(x)$ as,

\begin{widetext}
\begin{equation}
	\begin{bmatrix}
		v_F \tau_0 \otimes \left[ \bm{\pi}\vdot \bm{\sigma} + \frac{V(x)}{v_F} - \frac{\mu}{v_F} \right] && \Delta(T) \\
		\Delta^*(T) && - v_F \tau_0 \otimes \left[ \bm{\bar{\pi}}\vdot \bm{\sigma} + \frac{V(x)}{v_F} - \frac{\mu}{v_F} \right] 
	\end{bmatrix} \begin{bmatrix}
		\Psi_e\\
		\Psi_h
	\end{bmatrix}= \varepsilon \begin{bmatrix}
		\Psi_e\\
		\Psi_h
	\end{bmatrix} \label{GS}
\end{equation}
\end{widetext}

\begin{figure*}
	\centering
	\includegraphics[width=0.875 \linewidth]{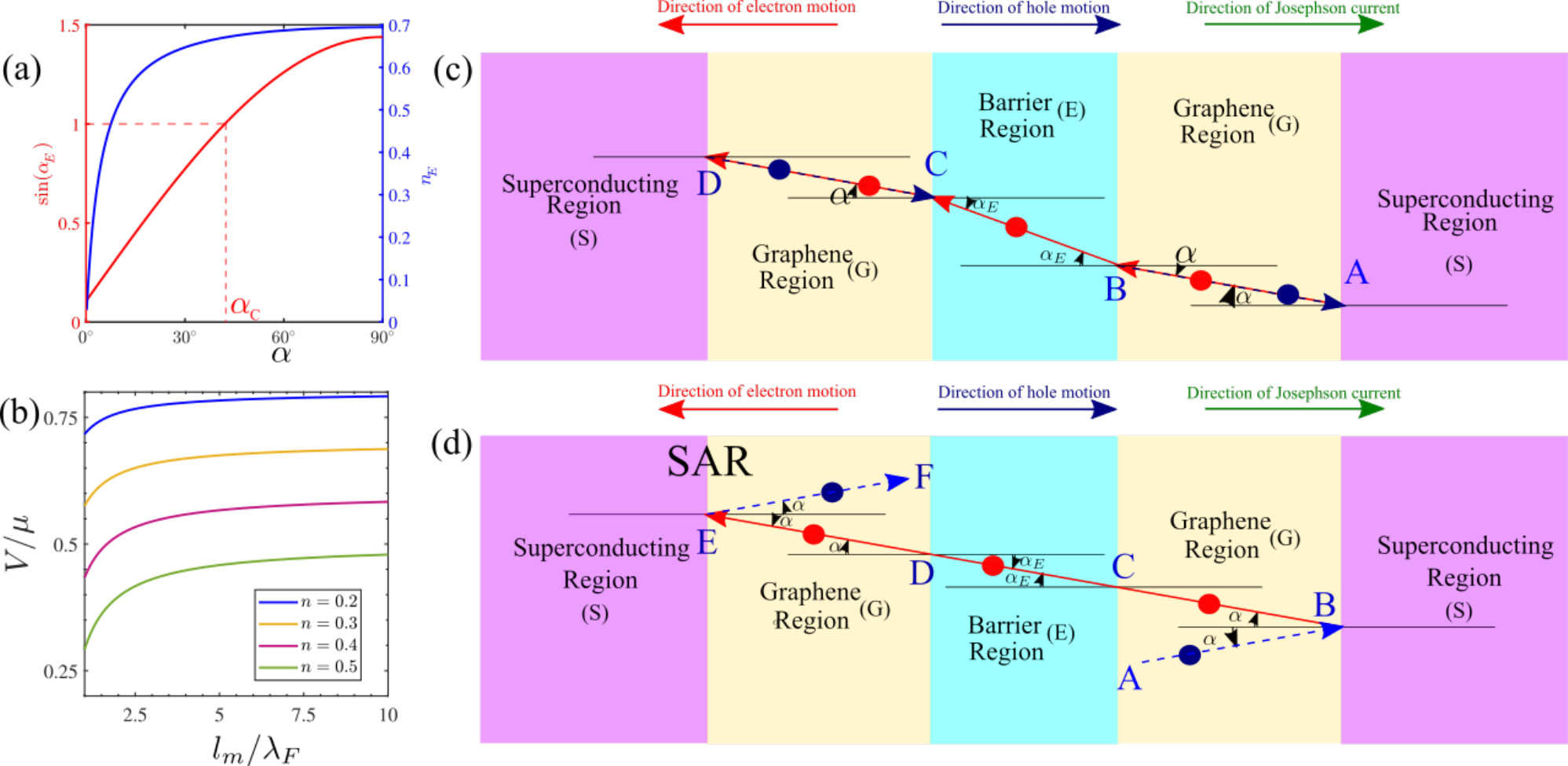}
	\caption{\justifying In (a) we show the plot of $\sin(\alpha_E)$ calculated from the Snell's law in the RAR ($\mu\gg \varepsilon$) regime for as a function of $\alpha$ using the red curve.  In the red curve we show the dependence of refractive index of the barrier region on $\alpha$ for a barrier with $\kappa_{lm}=0.5$ and $\kappa_V=0.25$.  In (b), we have plotted the values of barrier potential ($V$) and magnetic length ($l_m$) which share the given values of refractive index ($n_E$) for $\alpha=\pi/8$. Here $\mu$ and $\lambda_F$ is the Fermi energy and the Fermi wavelength of the barrier free region (G). In (c)  we show the propagation of electron and hole for the same barrier in the RAR regime. In the schematic diagram, a hole following path BA undergoes RAR in the GS interface at point A. Due to RAR, a reflected electron traces back the path of the incident hole AB with $\alpha=10^{\circ}$ and in the barrier region(E) acts as a medium with lighter refractive index ($n_E$) and the electron  goes through BC path with $\alpha_E=19.73^{\circ}$ and then again CD path through the G region with $\alpha_E=10^{\circ}$. After that at point D, the electron undergoes RAR in the GS boundary and again reflects back as a hole.  In (d), we show the schematic diagram for the propagation of electron and hole for the same barrier in the SAR regime. A hole from path AB with angle $\alpha'=10^{\circ}$ undergoes SAR in the GS interface at B and reflects back as an electron with $\alpha=10^{\circ}$. As the refractive index of the barrier region becomes $n_E \approx 1$ in the SAR regime, the reflected electron traces path BCDE and undergoes SAR and a reflected hole travels through EF direction. The blue solid circle denotes a hole and an electron is represented by a red solid circle.}
	\label{diag}
\end{figure*}
Where $\bm{\pi}=(\mathbf{p}+ e \mathbf{A}(x)) $ and $\bm{\bar{\pi}}=(\mathbf{p}- e \mathbf{A}(x)) $. 
The scalar potential $V(x)$ does not invert the sign, but the magnetic vector potential $\mathbf{A}$ changes the sign under the time reversal operation \cite{Sakurai2020}.  We need the solutions for the DBdG equation in our system for energies below the superconducting gap ($\Delta_0(T)$) of the superconductor.
\begin{comment}
	 Let us take s-wave superconductors for our present analysis where the pair potential in the superconductors is defined by,
	
	\begin{equation}
		\Delta=\begin{cases}
			\Delta_0 e^{i \phi_1} &\text{ at, } x<0\\
			0  &\text{ at, } 0<x<L \\
			\Delta_0 e^{i \phi_2} &\text{ at, } x>L \\
		\end{cases}
	\end{equation}.
\end{comment}

The Josephson current depends on the phase difference between the two superconducting regions ($\phi=\phi_2-\phi_1$). Here if we define the electron and hole excitations $\Psi_e$ and $\Psi_h$ as,
$\mqty[\psi_{e1} & \psi_{e2} & \psi_{e3} & \psi_{e4}]^T$ and $\mqty[\psi_{h1} & \psi_{h2} & \psi_{h3} & \psi_{h4}]^T$ then it is sufficient to calculate the solutions of $\psi_{e1}$, $\psi_{e2}$, $\psi_{h1}$ and $\psi_{h2}$ from \eqaref{DBDG} because of the basis we have chosen. From these solutions, the full $(8\times1 )$ wave function can be evaluated by 
\begin{equation}
	\Psi=\mqty[
		\mqty(
			C_{e1} \\ C_{e2}
		) \otimes \mqty(
			\psi_{e1}\\ \psi_{e2}
		) \\
		\mqty(
			C_{h1} \\ C_{h2}
		) \otimes \mqty(
			\psi_{h1}\\ \psi_{h2}
		)
	]
\end{equation}

Now the $(8 \times 8)$ DBdG equation transforms into two equivalent four-dimensional DBdG equation,

\begin{widetext}
	\begin{equation}
		\begin{bmatrix}
			V(x)- \mu && v_F (\mathbf{\pi}_x- i \mathbf{\pi}_y) &&  \Delta(T) && 0 \\
			v_F (\mathbf{\pi}_x+ i \mathbf{\pi}_y) && V(x)- \mu  && 0 &&  \Delta (T) \\
			\Delta^*(T)  && 0 && \mu - V(x) && - v_F (\mathbf{\bar{\pi}}_x-i \mathbf{\bar{\pi}}_y)\\
			0 && \Delta^*(T)  && - v_F (\mathbf{\bar{\pi}}_x+i \mathbf{\bar{\pi}}_y) && \mu -V(x)
		\end{bmatrix} \begin{bmatrix}
			\psi_{e1}\\ \psi_{e2}\\ \psi_{h1}\\ \psi_{h2}
		\end{bmatrix}= \varepsilon \begin{bmatrix}
			\psi_{e1}\\ \psi_{e2}\\ \psi_{h1}\\ \psi_{h2} 
		\end{bmatrix} \label{dbdg}
	\end{equation}
\end{widetext}

\begin{comment}
	\begin{figure*}
		\centering
		\includegraphics[width=\linewidth]{Plots.pdf}
		\caption{The Josephson Current is plotted with its dependence on (a) temperature, (b) strength of magnetic barrier ($\kappa_{lm}$) and (c) strength of barrier electrostatic potential ($\kappa_V$) and (d) ratio of size of EVMP regions and pure-graphene regions for an SG(EG)$^n$S type Josephson junction with $n=10$ in the RAR regime. In (e), (f) and (g) the Josephson current is plotted with temperature, $\kappa_{lm}$ and $r$ respectively for different values of $\phi$ again for the RAR regime. The Josephson current for the SAR regime is shown in (h). The temperature dependence of Josephson current in the case of SAR is shown in (i). The dependence of SAR current on $\kappa_{lm}$ is shown in (j). }
		\label{Current}
	\end{figure*}
\end{comment}

\begin{figure*}
	\centering
	\includegraphics[width=1.0\linewidth]{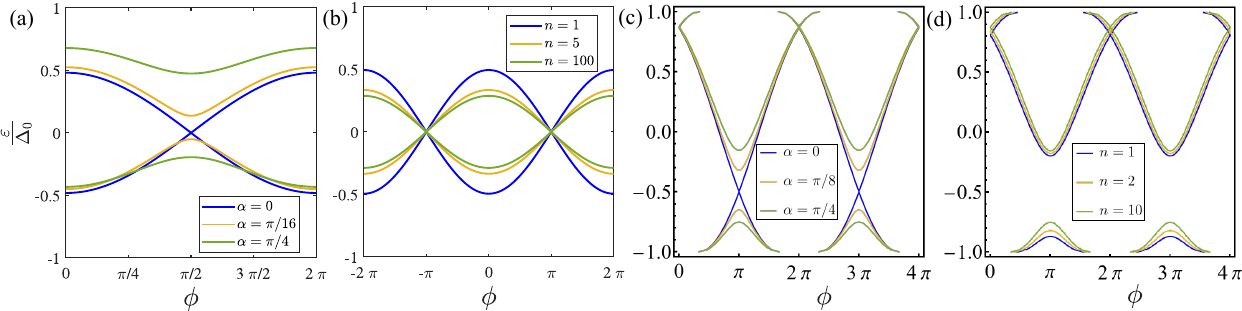}
	\caption{\justifying The $\varepsilon-\phi$  relation for SG(EG)$^n$S for retro and specular Andreev reflection is shown in (a),(b) and (c),(d) respectively. In (a) we show the $\varepsilon-\phi$  relation \color{black} for different values $\alpha$ for $\kappa=1$, $\kappa_V=0.5$, $n=10$ and $\kappa_{lm}=3$. In (b) the $\varepsilon-\phi$  relation is shown in case of RAR for different values of $n$ and for $\kappa=1$, $\kappa_V=0.5$, $\kappa_{lm}=3$ and $\alpha=\pi/4$. For the case of SAR, we show the $\varepsilon-\phi$  relation for different values of $\alpha$  in an SG(EG)$^n$S junction with $\kappa$=1, $\kappa_V=0.5$, $\kappa_{lm}=1$ and $n=10$. $\varepsilon-\phi$  relation is plotted in (d) for the case of SAR for different values of $n$ for $\kappa=1$, $\kappa_V=2$, $\kappa_{lm}=2$ and $\alpha=\pi/4$. }
	\label{ephi}
\end{figure*}

\section{Solutions of DBdG Equation and dispersion relation :}

\subsection{Snell's law}\label{Snell}
The two superconductors at the two ends of our system acts as Andreev reflectors for the graphene electrons as depicted in \figref{diag} (b) and (c). When an electron encounters the graphene (G)- superconductor(S) interface, it is reflected as a hole and a hole is reflected as an electron \cite{Andreev1965,Beenakker2006}. Between these two reflectors, the barriers (E) in the graphene region now act as mediums with modulated refractive index. 

The solutions of the DBdG equation in all the regions is discussed in detail in \apref{wf}. Using the properties of these wavefunctions, we can get Snell's law for graphene electrons propagating from the free region to the barrier region. 
%Let us define a few parameters which will be useful for further analysis,  $\kappa_V=\frac{V}{\mu}$, $\kappa_{lm}=\frac{\lambda_F}{l_m}$ and $\kappa=\frac{\mu d }{\hbar v_F}$. Here $\lambda_F$ is the Fermi wavelength in the graphene region of our system.
The direction of the propagation of electron and the hole can be obtained from their respective wavefunctions.
If the angle of propagation of graphene electron with respect to normal direction in free region and the barrier region is $\alpha$ and $\alpha_E$ respectively, then for the case of RAR ($\mu\gg \varepsilon$),  the electronic analogue of the refractive index in the G region  that captures 
the effect of the barriers in this region can be gives as 
\begin{equation}
	n_E=\frac{\sin(\alpha)}{\sin(\alpha_E)}=\frac{\left(1-\frac{V}{\mu}\right)k_F l_m \sin(\alpha)}{1+ k_F l_m \sin(\alpha)}. \label{sl}
\end{equation}
%From \eqaref{sl}, we can observe that such refractive index for the barriers in the case of RAR depends on the 
%\clr{
%incident angle ($\alpha$) and the height of both electrostatic and magnetic potential barrier.
%}
%strength of the electrostatic potential barrier ($\kappa_V$), strength of the magnetic barrier ($\kappa_{lm}$) and the incident angle ($\alpha$). 
In \figref{diag}(a) we have shown the value of $\sin(\alpha_E)$ calculated from the Snell's law defined in \eqaref{sl}.

For the case of RAR, the critical angle of incidence for graphene electron is,
\begin{equation}
	\alpha_C=\sin[-1](1-\frac{V}{\mu}-\frac{1}{k_F l_m}). \label{crit}
\end{equation}
Beyond the critical angle of incidence the propagating solution representing scattering states in the G region become evanescent waves ( bound states). Their hybridization with the Andreev bound states may lead to 
interesting features somewhat akin to the one studied in electron transport in SGS type of JJs in the presence of uniform magnetic to detect the valley polarisation of edge states produced in the graphene region\cite{Akhmerov2007}.
We shall not discuss this issue any further in the current work and will shelve it for future work. \eqaref{sl} that gives the electronics analogue of  Snell's law shows that different values of  electrostatic barrier potential (V) and magnetic length ($l_m$)  can give the same refractive index ($n_E$) for a given angle of propagation $\alpha$. which can be expressed as 
\begin{equation}
		V_2=\mu\left[1+\left(\frac{V_1}{\mu}-1\right)\right]\left(\frac{l_{m1}}{l_{m2}}\right) \left[\frac{1+ k_F l_{m2} \sin \alpha}{1+ k_F l_{m1}\sin \alpha}\right]. \label{gate}
\end{equation}
To elucidate this issue further in \figref{diag} (b), we have plotted the values of barrier potential ($V$) and magnetic length ($l_m$)  with which a given refractive index ($n_E$) can be achieved for a given angle of propagation ($\alpha$). 
The interconvertibility between the electric and magnetic field as they are applied to charge carriers in graphene that are massless Dirac fermions was studied extensively in various  contexts in a number of earlier works \cite{Lukose2007, Tan2010, Sharma2011}. 
In the current work we demonstrate its implication for JC through SGS junctions. 

%\begin{comment}
%We can also observe from this Snell's law that for a given value of refractive index ($n_E$), $\kappa_{lm}$ and angle of propagation ($\alpha$) the value of $\kappa_V$ is given by the straight line equation,
%
%\begin{equation}
%	\kappa_V=-\left[\left(\frac{n_E}{2 \pi}\right)\frac{1}{\sin\alpha}\right] \kappa_{lm} +(1-n_E).
%\end{equation}
%
%We have shown the values of $\kappa_{lm}$, $\kappa_V$ which share the same value of refractive indexes ($n_E$) for a given value of angle of propagation ($\alpha$) in \figref{diag} (c) .
%\end{comment}
	
In \figref{diag} (c) we have depicted a schematic diagram of motion of electron and hole in the RAR regime. In case of RAR, the reflected electron/hole traces back the path of incident hole/electron. The E region in this figure is the barrier region. It acts as a rarer medium compared to the G region for all possible angles of $\alpha$ as shown in \figref{diag} (a).
However, for the case of SAR ($\mu\ll \varepsilon$), we have, 

\begin{equation} \label{sarsl}
	n_E = \frac{\sin(\alpha)}{\sin(\alpha_E)} \approx 1.
\end{equation}
 After traveling through the barrier region with $\alpha<\alpha_C$, the charge carriers in graphene again moves with the same angle of propagation $\alpha$. The propagation of electrons and holes in the SAR regime is schematically depicted in \figref{diag} (d). In this case, the angle of reflection in the GS interface is same as angle of incidence, as happens in specular optical reflection. Again, when the graphene electron encounters the superconducting surface, only the electrons with angle of propagation($\alpha$) less than $\alpha_A=\sin[-1](|\varepsilon-\mu|/\varepsilon+\mu)$ undergoes Andreev reflection \cite{Beenakker2006}.
 
 % The schematic diagram of graphene electrons passing through the barrier region and undergoing Andreev reflection is shown in \figref{diag} (b) and (c) for RAR and SAR respectively. 

\begin{figure*}
	\centering
	\includegraphics[width=\linewidth]{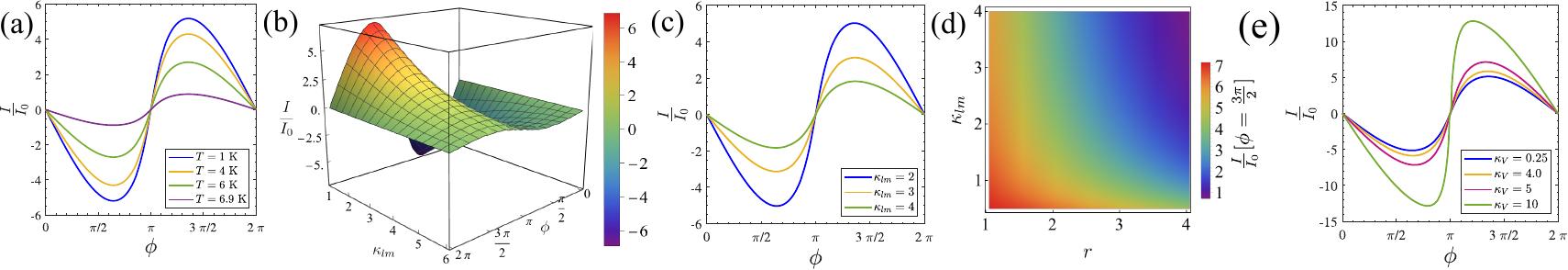}
	\caption{\justifying In (a) we show the $\phi$ and $T$ dependence of the Josephson
		current. For a constant temperature, the Josephson current is periodic in
		$\phi$ for an SG(EG)$^n$S type Josephson junction with $n=10$ in the RAR
		regime. In this case, we have taken, $\kappa_{lm}= 2.0 $ and $\kappa_V = 0.5$.  In (b) we show the Josephson Current as a function of $\phi$ and
		the strength of magnetic barrier ($\kappa_{lm}$). We can observe from the
		3D plot that for a constant value of $\phi$, the Josephson current
		decreases with the increasing $\kappa_{lm}$. In (c) we show the cross-sectional plots to highlight this behaviour.  In (d) we
		combine the effect of $\kappa_{lm}$ and the ratio of size of EVMP regions
		and pure-graphene regions for an SG(EG)$^n$S type Josephson junction in the
		same RAR regime. In (b), (c) and (d), we have taken, $\kappa_V=0.5$.  In (e) we plot the Josephson current for different values
		of $\kappa_V$ with $\phi$ for $\kappa_{lm}=2.0$. In all these cases we have fixed $\kappa=1$.  }
	\label{Current}
\end{figure*}

\subsection{Wavefunctions and boundary value conditions} \label{boundary}

From the DBdG equation \eqaref{dbdg}, we get four plane wave solutions for all the three type\clr{s } of regions, \ie superconducting region, graphene region both outside and inside the barrier regime. In \eqaref{dbdg} let us define $u= \mqty[ \psi_{e1} & \psi_{e2}]^T$ and $v= \mqty[ \psi_{h1} & \psi_{h2}]^T$. The solutions of the DBdG equation for a given value of $\varepsilon$ and $q$ are shown in \apref{wf}. Using the wave functions we can evaluate some boundary value conditions to make transfer matrices in this system. As we are considering the Andreev bound state spectrum \ie $\varepsilon<\Delta_0(T) $ at the GS interface, the electron and hole part of the wave-function of DBdG equation \eqaref{dbdg} is related by \cite{Titov2006}

\begin{subequations}
	\begin{gather} \label{sup}
		v (x=0)= e^{-i \phi_1 +i \beta\sigma_x} u(x=0) \\
		u(x=L) = e^{i \phi_2+i \beta\sigma_x} v(x=L)
	\end{gather}
\end{subequations}

The details about the boundary value condition is discussed in \apref{boun}.
$\beta$ in this case is defined by $\cos\beta=\varepsilon/\Delta_0(T) $ and the range of $\beta$ is $\left(- \frac{\pi}{2},\frac{\pi}{2}\right)$.

In the graphene region outside each barrier, the two ends of the region  (\ie $x=(n-1)D$ and $x=nD-d$) are related by \cite{Titov2006},
\begin{subequations}
	\begin{gather} \label{g}
		v\left[x=(n-1)D\right]		= M_2\left(\varepsilon,q\right) v\left[x=nD-d\right] \\
		u\left[x=nD-d\right]= M_1\left(\varepsilon,q\right)	u\left[x=(n-1)D\right]
	\end{gather}
\end{subequations}
Here, $M_1\left(\varepsilon,q\right)=\Lambda_1 e^{-ik (D-d) \sigma_z} \Lambda_1$ and $M_2\left(\varepsilon,q\right)=\Lambda_2 e^{-i k' (D-d) \sigma_z} \Lambda_2$. 
Also, $k$ and $k'$ is the wave vector in $x$ direction \ie the longitudinal wave vector for electron and hole excitations respectively,  $\Lambda_1$ and $\Lambda_2$ are $(2\times2)$ matrices given by:

\begin{subequations}
	\begin{gather}
		\Lambda_1=\Lambda_1^{-1}=\frac{1}{\sqrt{2 \cos \alpha}} \mqty[
		e^{-i\alpha/2}	&	e^{i \alpha/2}\\
		e^{i\alpha/2}	&	-e^{-i \alpha/2}
		]	\\
		\Lambda_2=\Lambda_2^{-1}=\frac{1}{\sqrt{2 \cos \alpha}} \mqty[
		e^{i\alpha'/2}	&	e^{-i \alpha'/2}\\
		e^{-i\alpha'/2}	&	-e^{i \alpha'/2}
		]	\\
		\text{with}, \sin(\alpha)=\frac{\hbar v_F q}{\varepsilon+ \mu}, \hspace{0.1\columnwidth} \sin(\alpha')=\frac{\hbar v_F q}{\varepsilon- \mu} \label{alpha}
	\end{gather}
\end{subequations}

The angle $\alpha, \alpha'$ $\left(0, \frac{\pi}{2}\right)$ are the angles of incidence and reflection of the electrons and holes respectively in the NS interface. Similarly, inside the barrier regions in the graphene region, the wave functions in the two ends of $n$'th barrier of width $d'$ (\ie $x=nD-d$ and $x=nD$) are related by,
\begin{subequations}
	\begin{gather} \label{emvp}
		v(x=nD-d) = M_{2E}\left(\varepsilon,q\right) v(x=nD) \\
		u(x=nD) = M_{1E}\left(\varepsilon,q\right)	u(x=nD-d)
	\end{gather}
\end{subequations}

Here $M_{1E}\left(\varepsilon,q\right)=\Lambda_{1E} e^{-ik_E d \sigma_z} \Lambda_{1E}$ and $M_{2E}\left(\varepsilon,q\right)= \Lambda_{2E} e^{-i k'_E d \sigma_z} \Lambda_{2E} $. 
Also, $k_E$ and $k'_E$ are the wave vectors in the $x$ direction for electron and hole excitations inside the barrier, and $\Lambda_{1E}$ and $\Lambda_{2E}$ are $(2\times2)$ matrices.

\begin{subequations}
	\begin{gather}
		\Lambda_{1E}=\Lambda_{1E}^{-1}=\frac{1}{\sqrt{2 \cos \alpha_E}} \mqty[	e^{-i \alpha_E/2}		&		e^{i \alpha_E/2}\\
		e^{i \alpha_E/2}		&		-e^{-i \alpha_E/2}]\\
		\Lambda_{2E}=\Lambda_{2E}^{-1}=\frac{1}{\sqrt{2 \cos \alpha'_E}} \mqty[	e^{i \alpha'_E/2}		&		e^{-i \alpha'_E/2}\\
		e^{-i \alpha'_E/2}		&		-e^{i \alpha'_E/2}] \\
		\sin(\alpha_E)=		\left(\frac{\hbar v_F	\left(q + \frac{1}{l_m}\right)}{\varepsilon -V + \mu}\right), \hspace{0.01\columnwidth} \sin(\alpha'_E)=		\left(\frac{\hbar v_F	\left(q - \frac{1}{l_m}\right)}{\varepsilon -\mu + V}\right)
	\end{gather}
\end{subequations}

%Let us also define the ratio of the width of the EMVP barrier region and the normal region as $\Delta$.

\subsection{ Dispersion  $(\varepsilon-\phi)$relation} \label{disp}
Let us consider $n$ electrostatic and magnetic vector potential (EMVP) barriers of width $d$ in the graphene region of the SGS junction separated by a distance $D$ such that $L=ND + (D-d)$. The condition for a bound state in this system is that the transfer matrix for a round trip from $x=0$ to $x=L$ and again from $x=L$ to $x=0$ to be a unit matrix of dimension $(2\times2)$ \cite{Titov2006}.  Using \eqaref{sup}, \eqref{g} and \eqref{emvp} it can be shown that this condition results in,

\begin{subequations}
	\begin{align}
		&Det\left[\mathbb{I}_2- e^{i\phi} T (\varepsilon,q)\right]=0 \label{tr} \\
		T (\varepsilon,q)=& M_1(\varepsilon,q) \left[	M_{1E}(\varepsilon,q)M_{1}(\varepsilon,q)	\right]^n e^{i \beta \sigma_x}\notag \\
		& \hspace{0.1 \columnwidth}\left[	M_{2}(\varepsilon,q) M_{2E}(\varepsilon,q)	\right]^n M_{2}(\varepsilon,q) e^{i \beta \sigma_x} 
	\end{align}
\end{subequations}

This equation can be solved for any value of $\varepsilon$, $q$ and $\phi$. The $\varepsilon-\phi$ relation calculated from \eqaref{tr} can be written as an equation for a conic section \cite{Salim2023},

 \begin{figure*}
	\centering
	\includegraphics[width=\linewidth]{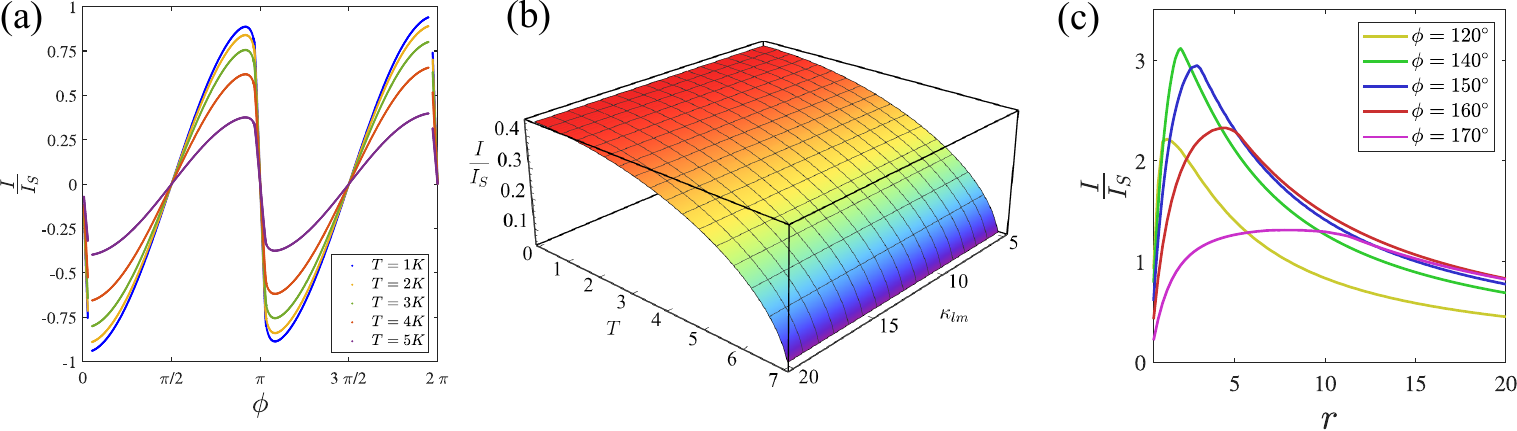}
	\caption{\justifying The Josephson Current is plotted with its dependence on $\phi$ for an SG(EG)$^n$S type Josephson junction with $n=10$ in the SAR regime for different values of temperature \clr{ while keeping $\kappa_{lm}=10$ and $\kappa_V=2$. } In (b) we show the Josephson current for different values of $\kappa_{lm}$ and temperature ($T$) for $\phi=2 \pi /3$ \clr{ and $\kappa_V=2$ }. We can observe that as we are in the short junction limit, the $\kappa_{lm}$ does not impact the Josephson current. In (c) the Josephson current is plotted with $r$ for different values of $\phi$ again for the SAR regime \clr{ while keeping $\kappa_{lm}=10$ and $\kappa_V=2$. In all these cases we have taken $\kappa$=1}.}
	\label{CurrentSAR}
\end{figure*}

\begin{equation}
	A-B y^2 + C xy+ \cos(\phi)=0
\end{equation}
where $A$, $B$, $C$ and $D$ are constants which depend on the parameters defining and characterising the barrier regime (such as $\kappa_V=\frac{V}{\mu}$, $\kappa_{lm}=\frac{\lambda_F}{l_m}$, $\kappa=\frac{\mu d }{\hbar v_F}$ and the angles $\alpha$ and $\alpha'$), $x=\sin(\varepsilon/\Delta_0)$ and $y=\cos(\varepsilon/\Delta_0)$. Here $\lambda_F$ is the Fermi wavelength in the barrier free graphene region (G) of our system. There are two regimes of energy which are interesting to study, retro Andreev reflection (RAR) and specular Andreev reflection (SAR). RAR dominates in the superconductor-graphene junctions when $\mu\gg\Delta_0(T) $\cite{Beenakker2006}. From \eqaref{alpha}, we can see that in this regime $\alpha'=-\alpha$. In the regime of $\mu\ll \Delta_0(T) $, the SAR dominates. From \eqaref{alpha}, we can show that in the SAR regime $\alpha'=\alpha$.
% We have shown the \color{blue} $\varepsilon-\phi$ relation \color{black} in \figref{ephi}. 
In \figref{ephi} (a) and (b) we show the $\varepsilon-\phi$ relation for the RAR regime and in  (c) and (d) we show the $\varepsilon-\phi$ relation for the SAR regime. As a prototype case, for $10$ such barriers inside the graphene region of the SGS junction, we can observe that the $\phi$ dependence on $\varepsilon$  for $\alpha\ne0$, is asymmetric with respect to $\varepsilon=0$ in our system. This is not the case in normal SGS junctions. For $\alpha\ne0$, the gap opens at $\phi=0$. In (b) we show the dependence of the $\varepsilon-\phi$ relation on the number of barriers we have considered in the graphene region. As we increase the number of barriers, the two values of energy for a fixed value of $\phi$ come closer. In (c) we show the $\varepsilon-\phi$ relation for the SAR regime. Again, similar to RAR we can see for $\alpha=0$ the gap closes. Again in (d) we show the dependence of the number of barriers for the $\varepsilon-\phi$ relation in the SAR regime. As we see in the next section \ref{curr}, this change in the dispersion strongly influences the Josephson current in the presence of such barriers and provides us a way to tune the junction properties.

\section{Josephson Current} \label{curr}

Josephson current $I$ across this junction at a temperature $T$ is given by \cite{Golubov2004,Titov2006, Maity2007, zagoskin2012} 

\begin{gather}
	I(\phi, T)= \frac{4 e}{\hbar} \sum_n \sum_{q=-k_F}^{k_F} \left( \pdv{\varepsilon_n}{\phi}\right) f(\varepsilon_n),
\end{gather}
where $f(\varepsilon_n)$ is the Fermi distribution function. In \figref{Current} (a) we plot the Josephson current as a function of $\phi$ for different temperatures  in the RAR regime.
In (b) and (c), we show the effect of strength of magnetic barrier on the Josephson current using a 3D plot and 2D plot respectively in the RAR regime. The magnetic barrier has a pronounced effect on the magnitude of the Josephson current. 
We define the ratio of the length of barrier region (E) and the barrier free graphene region (G) as $r$. By combining the effect both this ratio $r$ and the strength of the magnetic field ($\kappa_{lm}$) we show in (d) how the Josephson current can be tuned. In (e) we show the effect of $\kappa_V$, which denotes the strength of the electrostatic scalar potential, on the Josephson current.
Here, $I_0=\frac{8e k_F \Delta_0(T=0)}{\hbar}$ and for demonstrative purpose in all of these cases, we have considered $10$ EMVP barriers inside the graphene region of the SGS junction. 

For the SAR regime, we show the Josephson current in \figref{CurrentSAR} (a) for different temperatures. In the short junction limit, the SAR current does not depend on the strength of the magnetic barrier ($\kappa_{lm}$), unlike the RAR case which is shown in (b). This happens because the transfer matrices $M_{1E}$ and $M_{2E}$ in \eqaref{tr} become unit matrices in this limit. Also, we can observe from the Snell's law in the SAR regime \eqaref{sarsl} that the refractive index of the barriers approaches unit value in the SAR regime, thereby  making the magnetic barrier transparent to the Josephson current in the SAR regime. 
 
 To conclude we have shown that the Josephson current in a SGS junction can be significantly tuned by magnetically modulate the graphene regime.
 Additionally the strength of this magnetic barrier can be tuned by introducing gate voltage. This combination thus provides an important tuning parameter to control Josephson current through such junctions and indicates possibility of interesting device applicability.
 Following the work on electron transport in SGS type of JJs in the presence of uniform magnetic field that shows hybridisation of edge states in the graphene region with andreev bound state \cite{Akhmerov2007}, superposition of the uniform magnetic field and the magnetic barrier in the graphene region \cite{PhysRevB.98.125303} are expected to 
 show rich physics with wider possibility of device application. This could be an interesting topic for future research in this direction.
 
 PSB is supported by a MHRD Fellowship.
 SG and RM are supported by a SPARC Phase II (MHRD, GOI, Project Code P2117)  grant, PSB and SG are also partially supported by a MFIRP Project (MI02545) of IIT Delhi.

  \appendix
 
 \section{Wavefunctions in the Three Regions} \label{wf}
 
The wave-functions in the three types of region considered can be calculated for a given value of $\varepsilon$, $q$ and $\phi$ from \eqaref{dbdg}. 

\subsection{Superconducting Region}

In the superconducting region, the DBdG equation from \eqaref{dbdg} is given by,

\begin{widetext}
	\begin{equation}
		\begin{bmatrix}
			- \mu -U_0 && v_F (p_x- i p_y) &&  \Delta_0 e^{i \Phi} && 0 \\
			v_F (p_x+ i p_y) && - \mu -U_0 && 0 &&  \Delta_0 e^{i \Phi} \\
			\Delta_0 e^{-i \Phi} && 0 && \mu +U_0  && - v_F (p_x-i p_y)\\
			0 && \Delta_0 e^{-i \Phi} && - v_F (p_x+i p_y) && \mu  +U_0
		\end{bmatrix} \begin{bmatrix}
			\psi_{e1}\\ \psi_{e2}\\ \psi_{h1}\\ \psi_{h2}
		\end{bmatrix}= \varepsilon \begin{bmatrix}
			\psi_{e1}\\ \psi_{e2}\\ \psi_{h1}\\ \psi_{h2} 
		\end{bmatrix} \label{supl}
	\end{equation}
\end{widetext}

Here, $\Phi$ is the superconducting phase. The possible solutions in the regime of $U_0+\mu \gg \Delta_0, \varepsilon$ are given by

\begin{subequations} \label{super}
	 \begin{equation}
		\psi_1 (\Phi)=e^{iqy+ik_ox-k_i x}\left(
		\begin{array}{ccc}
			e^{i\beta} \\ 	
			e^{i\beta+i\gamma} \\ 
			e^{-i\Phi} \\
			e^{-i\Phi+i\gamma} \\
		\end{array} 
		\right),
	\end{equation}
	\begin{equation}
		\psi_2 (\Phi)=e^{iqy+ik_ox+k_i x}\left(
		\begin{array}{ccc}
			e^{-i\beta} \\ 	
			e^{-i\beta+i\gamma} \\ 
			e^{-i\Phi} \\
			e^{-i\Phi+i\gamma} \\
		\end{array} 
		\right),
	\end{equation}
	\begin{equation}
		\psi_3 (\Phi)=e^{iqy-ik_ox-k_i x}\left(
		\begin{array}{ccc}
			e^{-i\beta} \\ 	
			-e^{-i\beta-i\gamma} \\ 
			e^{-i\Phi} \\
			-e^{-i\Phi-i\gamma} \\
		\end{array} 
		\right),
	\end{equation}
	\begin{equation}
		\psi_4 (\Phi)=e^{iqy-ik_ox+k_i x}\left(
		\begin{array}{ccc}
			e^{i\beta} \\ 	
			-e^{i\beta-i\gamma} \\ 
			e^{-i\Phi} \\
			-e^{-i\Phi-i\gamma} \\
		\end{array} 
		\right).
	\end{equation}
\end{subequations}

where, 

\begin{equation} \label{16}
	\beta =
	\left\{
	\begin{array}{ll}
		\cos^{-1}\left( \frac{\varepsilon}{\Delta_o} \right)  & \mbox{if } \varepsilon < \Delta_o \\
		-i \cosh^{-1}\left( \frac{\varepsilon}{\Delta_o} \right) & \mbox{if } \varepsilon > \Delta_o
	\end{array}
	\right.
\end{equation}

\begin{equation} \label{51}
	\gamma= \sin^{-1}\left[ \frac{\hbar v_F q}{U_o+\mu} \right]
\end{equation}

\begin{equation} \label{52}
	k_o=\sqrt{\left(\frac{U_o+\mu}{\hbar v_F}\right)^2 -q^2}
\end{equation}
\begin{equation} \label{53}
	k_i=\frac{(U_o+\mu)\Delta_o}{\hbar^2 v_F^2 k_o}\sin \beta
\end{equation}

In the regime of $|q|\le \frac{\mu}{\hbar v_F}$ and if we take $U_o\gg \mu ,\varepsilon$ then, $\gamma \rightarrow 0$, $k_o \rightarrow \frac{U_o}{\hbar v_F}$ and $k_i \rightarrow \frac{\Delta_o}{\hbar v_F} \sin \beta $. In the region $x<0$, (\ie left superconductor), the wavefunction is $\Psi_l= a_1 \psi_2 (\phi_1) + a_2 \psi_4 (\phi_1) $. In the region $x>L$, (\ie right superconductor) the wave-function is $\Psi_{r}= b_1 \psi_1(\phi_2) +	b_2 \psi_3(\phi_2)$.

\subsection{Outside the EMVP Barrier}

Outside the EMVP barrier in the graphene region, the wave-functions are given by,

\begin{subequations} \label{gra}
	\begin{equation}\label{85}
		\mathbf{\psi^{e+}}=\frac{e^{iqy+ikx}}{\sqrt{\cos \alpha}}\left(
		\begin{array}{ccc}
			e^{-\frac{i \alpha}{2}} \\ 	
			e^{\frac{i \alpha}{2}} \\ 
			0 \\
			0 \\
		\end{array} 
		\right),
	\end{equation}
	
	\begin{equation}\label{86}
		\mathbf{\psi^{e-}}=\frac{e^{iqy-ikx}}{\sqrt{\cos \alpha}}\left(
		\begin{array}{ccc}
			e^{\frac{i \alpha}{2}} \\ 	
			-e^{-\frac{i \alpha}{2}} \\ 
			0 \\
			0 \\
		\end{array} 
		\right),
	\end{equation}
	
	\begin{equation}\label{87}
		\mathbf{\psi^{h+}}=\frac{e^{iqy+ik'x}}{\sqrt{\cos \alpha'}}\left(
		\begin{array}{ccc}
			0 \\ 	
			0 \\ 
			e^{-\frac{i \alpha'}{2}} \\
			-e^{\frac{i \alpha'}{2}} \\
		\end{array} 
		\right),
	\end{equation}
	
	\begin{equation}\label{88}
		\mathbf{\psi^{h-}}=\frac{e^{iqy-ik'x}}{\sqrt{\cos \alpha'}}\left(
		\begin{array}{ccc}
			0 \\ 	
			0 \\ 
			e^{\frac{i \alpha'}{2}} \\
			e^{-\frac{i \alpha'}{2}} \\
		\end{array} 
		\right).
	\end{equation}
\end{subequations}

Where,

\begin{center}
	$\alpha=$ angle of incidence of electron $= \sin^{-1}\left[ \frac{\hbar v_F q}{\varepsilon + \mu}  \right]$ ,
\end{center}

\begin{center}
	$\alpha'=$ angle of reflection of hole$= \sin^{-1}\left[ \frac{\hbar v_F q}{\varepsilon - \mu}  \right]$ ,
\end{center}

\begin{center}
	$q=$ transverse wave vector,
\end{center}

\begin{center}
	$k=$ longitudinal wave vector of electron $= \left(\frac{\varepsilon + \mu}{\hbar v_F}\right) \cos \alpha$ ,
\end{center}

\begin{center}
	$k'= $ longitudinal wave vector of hole $= \left(\frac{\varepsilon - \mu}{\hbar v_F}\right) \cos \alpha'$ .
\end{center}

The wavefunction in this region is a linear superposition of all four basis states in \eqaref{gra}

\subsection{Inside the EMVP barrier}

Inside the EMVP barrier in the graphene region, the wave-functions are given by,

\begin{subequations}\label{em}
	\begin{equation}\label{d11}
		\mathbf{\psi_E^{e+}}=\frac{e^{iqy+ik_Ex}}{\sqrt{\cos \alpha_E}}\left(
		\begin{array}{ccc}
			e^{-\frac{i \alpha_E}{2}} \\ 	
			e^{\frac{i \alpha_E}{2}} \\ 
			0 \\
			0 \\
		\end{array} 
		\right),
	\end{equation}
	
	\begin{equation}\label{d12}
		\mathbf{\psi_E^{e-}}=\frac{e^{iqy-ik_Ex}}{\sqrt{\cos \alpha_E}}\left(
		\begin{array}{ccc}
			e^{\frac{i \alpha_E}{2}} \\ 	
			-e^{-\frac{i \alpha_E}{2}} \\ 
			0 \\
			0 \\
		\end{array} 
		\right),
	\end{equation}
	
	\begin{equation}\label{d13}
		\mathbf{\psi_E^{h+}}=\frac{e^{iqy+i{k'}_E x}}{\sqrt{\cos \alpha'_E}}\left(
		\begin{array}{ccc}
			0 \\ 	
			0 \\ 
			e^{-\frac{i \alpha'_E}{2}} \\
			-e^{\frac{i \alpha'_E}{2}} \\
		\end{array} 
		\right),
	\end{equation}
	
	\begin{equation}\label{d14}
		\mathbf{\psi_E^{h-}}=\frac{e^{iqy-ik'_E x}}{\sqrt{\cos \alpha'_E}}\left(
		\begin{array}{ccc}
			0 \\ 	
			0 \\ 
			e^{\frac{i \alpha'_E}{2}} \\
			e^{-\frac{i \alpha'_E}{2}} \\
		\end{array} 
		\right).
	\end{equation}
\end{subequations}

Here, 

\begin{subequations}
	\begin{equation}
		\sin(\alpha_E)=		\left(\frac{\hbar v_F	\left(q + \frac{1}{l_m}\right)}{\varepsilon -V + \mu}\right)
	\end{equation}
	\begin{equation}
		\cos(\alpha_E)=		\left(\frac{\hbar v_F	k_E}{\varepsilon -V + \mu}\right)
	\end{equation}
	\begin{equation}
		\sin(\alpha'_E)=		\left(\frac{\hbar v_F	\left(q - \frac{1}{l_m}\right)}{\varepsilon -\mu + V}\right)
	\end{equation}
	\begin{equation}
		\cos(\alpha'_E)=		\left(\frac{\hbar v_F	k'_E}{\varepsilon -\mu + V}\right)	
	\end{equation}
\end{subequations}

Here also the wave-function in this region is a linear superposition of all four basis states in \eqaref{em}.

\section{Boundary Value Conditions and Transfer Matrices} \label{boun}

The electron and hole part of the solutions of the DBdG equation \eqaref{dbdg} in the superconducting region are related by the transfer matrices $e^{-i \phi_1 +i \beta\sigma_x}$ and $e^{i \phi_2 +i \beta\sigma_x}$. In the left superconductor($x>L$), the only solutions possible from the four solutions in \eqaref{super} are $\Psi_l= a_1 \psi_2 (\phi_1) + a_2 \psi_4 (\phi_1) $. In the region $x>L$, (\ie right superconductor) the wave-function is $\Psi_{r}= b_1 \psi_1(\phi_2) +	b_2 \psi_3(\phi_2)$. At $x=0$ and $x=L$, the electron part $u$ and the hole part $v$ of these two wavefunctions $\Psi_l$ and $\Psi_r$ are related by \cite{Titov2006},

\begin{subequations}
	\begin{gather}
		v (x=0)= e^{-i \phi_1 +i \beta\sigma_x} u(x=0) \\
		u(x=L) = e^{i \phi_2+i \beta\sigma_x} v(x=L)
	\end{gather}
\end{subequations}

where $e^{i \beta \sigma_x}= \mqty[\cos(\beta)	&	i\sin(\beta)\\
i\sin(\beta)	&	\cos(\beta)]$ and $\phi_{1,2}$ are the phases of superconductors.

%\bibliography{References/reference.bib}
%apsrev4-2.bst 2019-01-14 (MD) hand-edited version of apsrev4-1.bst
%Control: key (0)
%Control: author (8) initials jnrlst
%Control: editor formatted (1) identically to author
%Control: production of article title (0) allowed
%Control: page (0) single
%Control: year (1) truncated
%Control: production of eprint (0) enabled
%
\end{document}